\documentclass[aps,pra,twocolumn,showpacs,floatfix,preprintnumbers,amsmath,amssymb]{revtex4}
\usepackage{graphicx}
\usepackage{dcolumn}
\usepackage{bm}

\begin{document}
\title{Entropic uncertainty relations for the ground state of a coupled system}
\author{M. S. Santhanam}
\affiliation{Max Planck Institute for the Physics of Complex Systems,\\
N\"othnitzer Strasse 38., Dresden 01187,\\ Germany}
\date{\today}

\begin{abstract}

There is a renewed interest in the uncertainty principle, reformulated
from the information theoretic point of view, called the entropic
uncertainty relations. They have been studied for various integrable
systems as a function of their quantum numbers. In this work,
focussing on the ground state of a nonlinear, coupled Hamiltonian system, 
we show that approximate eigenstates can be can be constructed within the
framework of adiabatic theory. Using the adiabatic eigenstates, we estimate the
information entropies and their sum as a function of the the nonlinearity
parameter. We also briefly look at the information entropies for the highly
excited states in the system.

\end{abstract}

\maketitle

\section{Introduction}
The uncertainty relations, that express the inability to simultaneously
measure the states of two non-commuting observables, form the cornerstone
of quantum physics. For any pair of operators $\widehat{A}$ and 
$\widehat{B}$, in standard form \cite{robert}, they are stated as,
\begin{equation}
\Delta A ~\Delta B \ge  \frac{1}{2} 
\langle \Psi|[\widehat{A},\widehat{B}]|\Psi \rangle
\label{stdur}
\end{equation}
where $\Delta A$ and $\Delta B$ represent the dispersions in
$\widehat{A}$ and $\widehat{B}$ and $[\widehat{A},\widehat{B}]$ is the
commutator.
In recent years, there has been a revival of interest in
the uncertainty relations reformulated from the stand point of
information theory, called the entropic uncertainty relations (EUR) \cite{eur1}.
For instance, the position-momentum uncertainty relation is formulated as follows;
Given an eigenstate of a quantum system, $\psi(q)$ and $\widetilde{\psi}(p)$, in
configuration and momentum space representations, and if $S_q$
and $S_p$ represent their information entropies, then the
entropic uncertainty relations can be written down as,
\begin{equation}
S_q + S_p \ge S_{qp}
\label{eur}
\end{equation}
where $S_{qp}$ is the lower bound to the entropic sum or EUR.
Here, the information entropy is defined as,
\begin{equation}
S_{\xi} = - \int^{\infty}_{-\infty} \rho(\vec{\xi}) \ln \rho(\vec{\xi}) ~d\vec{\xi}
\end{equation}
where $\rho(\vec{\xi}) = |\psi(\vec{\xi})|^2$ is the probability density.
The information entropy is a measure of the spreading or 
localisation of the given eigenstate.
Apart from its intrinsic value, the reformulation also seeks to
address some of the shortcomings in the standard statement of
the uncertainty principle \cite{eur1}. It also quantifies the uncertainty
more accurately than the standard statement based on dispersions \cite{bet}.
In particular, a good amount of work has focussed on obtaining
the lower bounds $S_{qp}$ in general and we mention the
result due to Bialynicki-Birula and Mycielski \cite{bbm},
\begin{equation}
S_q + S_p \ge D (1 + \ln \pi)
\label{bbmeq}
\end{equation}
where $D$ is the dimensionality of the system.
Inspired by this general result, several authors have focussed on
obtaining the lower bounds for EUR of quantum systems whose
classical limit is integrable, like the particle in an
infinite well \cite{infw}, harmonic oscillator and hydrogen atom 
in one and higher
dimensions \cite{ho}, power-law wavepackets \cite{pow} and
oscillating circular membrane \cite{cir}.
Apart from the lower bounds, the values of EUR as
a function of quantum numbers has been analysed in detail in
these series of papers. Recently, Dehesa et. al. have shown that
for the one-dimensional power-law potentials of the form $V(x) = x^{2k}$,
where $k$ is a positive integer, in the region of highly excited states, the
entropic sum goes as $\ln(2n)$ for all $k$ \cite{deh}.

On the other hand, the single-particle probability density $\rho(\vec{\xi})$
is also the quantity of fundamental interest in the density functional
theory and hence its characterisation using information entropy as
a measure for spreading has assumed
special interest. In fact, treating atomic and molecular entropic sum was
considered by Gadre \cite{gad1} and he derived
an approximate expression for the entropic sum within the Thomas-Fermi
framework for neutral atoms.
Now it is known from several empirical studies for atomic,
molecular and nuclear distributions that the entropic sum can be modelled
as \cite{amnc},
\begin{equation}
S \equiv c_1 + c_2 \ln N
\label{geq}
\end{equation}
where $c_1$ and $c_2$ are constants and $N$ is number of electrons or nucleons,
as the case maybe.
The functional form given above seems to be fairly universal
for many-fermions in some mean interactions \cite{amnc}.

Thus, one branch of the work on EUR has focussed
on quantum systems in the classically integrable limit whose eigenstates
are analytically known. The other complementary branch has explored the
complex atomic and molecular systems using a combination of
approximate analytical and empirical methods. In these cases, the EURs
have been obtained as a function of increasing quantum numbers. However, simple
and chaotic model systems that bridge this divide between the
purely integrable and the complex many-body systems have not yet been
considered. The main reason seems to be that, as yet, no straightforward
analytical technique is available to determine their eigenstates.
Thus, the main purpose of this article is to show that using adiabatic technique
approximate eigenstates can be constructed and further that the entropic sums
can also be estimated.
In contrast to earlier works,
we study the EUR as a function of the nonlinearity parameter in the system. Thus, we hope
to understand the effect of nonlinearity using information theoretic measures.

In general, the non-integrable Hamiltonian systems are generic rather than
an exception and model realistic physical systems.
Their phase space presents a mixture of regular and irregular trajectories,
whose properties change with a parameter. The evolution
of the probability density as a function of the parameter reflects
the influence of these classical objects.
Thus it is important to look at the EUR upon variation
of the nonlinearity
parameter. Note that scale invariance of entropic sums leads to
EURs that are independent of scaling parameters in an integrable
system. But, in a non-integrable system, the entropic
sums could depend on the nonlinearity parameter.

\section{Model Hamiltonian}

\begin{figure}
\includegraphics[height=5cm]{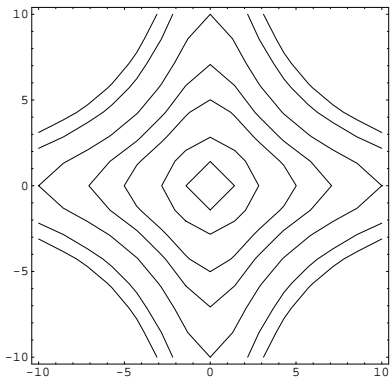}
\caption{The contours of the potential as a function of $x$ and $y$
for $\alpha=0.2$ and $k_1, k_2=1$. Each contour represents a particular
value of energy. For large values of energy, the potential develops 'channels',
as seen in the figure.}
\label{potential}
\end{figure}

In this work, we will consider the model Hamiltonian,
\begin{equation}
H = \frac{p_x^2}{2} + \frac{p_y^2}{2} + \frac{k_1^2 x^2}{2} + \frac{k_2^2 y^2}{2} + \alpha x^2 y^2
\label{Ham}
\end{equation}
whose potential is displayed in Fig. \ref{potential} and
$k_1, k_2, \alpha$ are the parameters. The system is integrable
for $\alpha=0$
and corresponds to two-dimensional harmonic oscillator whose
asymptotic (large quantum number) entropic sum was recently derived \cite{ho}. 
For a choice of $\alpha=0.05$ ($k_1=k_2=1$), the system displays
classical chaos for energy $E>15$ \cite{cho}.
The coupled oscillators are popular as models of chaos since they qualitatively
mimic the complex dynamics in
problems like the hydrogen atom in strong fields and generalised
van der Waals potential \cite{ganesan}.
For a generic chaotic system, most
of its eigenstates could be thought of as random waves except for a small sub-set
of 'regular' states that are influenced by the nature of local classical dynamics.
Hence in this work, we will only focus
attention, for most part, on the ground state and, briefly,
on the regular states. We stress that even though the system is
chaotic, we study only those eigenstates that are associated locally with regular
classical structures. Ground state is pre-eminent because
previous studies have shown that the ground state saturates the EUR inequality
in some systems (e.g. harmonic oscillator) and often in complex systems, the
focus is on the ground state properties.
Next, we obtain the ground state entropies, as a function of $\alpha$.

\section{Adiabatic Theory}
In the context of two degrees of freedom systems, if the frequency
of oscillations between both the degrees of freedom differ vastly,
then such
a classical system becomes an ideal candidate for the adiabatic treatment.
This is the well known Born-Oppenheimer approximation in atomic physics \cite{boa}.
The approach is to average over the 'faster' degree of freedom and incorporate
it into the Hamiltonian for the slower degree of freedom.
In this adiabatic approach, the Hamiltonian is effectively integrable
and can be quantised. This has been successfully
implemented earlier to estimate the energies of the localised states in coupled
oscillator systems \cite{adia,cho}. For the Hamiltonian in Eq. (\ref{Ham}), 
Certain and Moiseyev construct adiabatic eigenstates formally \cite{cho}.
Prange et. al. construct adiabatic eigenstates for a class of chaotic billiards
\cite{prange}.

\subsection{Position Eigenstates}
In this section, we will obtain the ground state of the Hamiltonian in Eq. (\ref{Ham}),
for $\alpha << 1$, in the adiabatic approximation, closely following the
technique in Refs. \cite{cho,adia}. We assume that the $y$-motion is
faster and average over the fast
motion by dropping terms that depend purely on $x$ and $p_x$. This gives,
\begin{equation}
H'(x;y) = \frac{p_y^2}{2} + \frac{k_2^2 y^2}{2} + \alpha x^2 y^2 = E'(y)
\end{equation}
where the variable $x$ is treated as a parameter. 
This is just the harmonic oscillator whose frequency is
\begin{equation}
\Omega_x = \sqrt{k_2^2 + 2\alpha x^2}
\label{full_omega}
\end{equation}
and for $\alpha << 1$, such that $k_2^2 >> 2\alpha x^2$, then
\begin{equation}
\Omega_x \backsimeq \omega_x = k_2 + \frac{\alpha}{k_2} x^2
\label{app_omega}
\end{equation}
The ground state of $H'(x;y)$ for $\alpha << 1$ is,
\begin{equation}
\phi(x;y) = \left( \frac{\omega_x}{\pi} \right)^{1/4} \exp(- \omega_x y^2/2 )
\end{equation}
Now, the adiabatic Hamiltonian is obtained as,
\begin{equation}
H_{ad} = \frac{p_x^2}{2} + \frac{k_1^2 x^2}{2} + \omega_x J_y
\end{equation}
where $J_y$ is the classical action corresponding to the faster motion.
In terms of action-angle coordinates, we get,
\begin{equation}
H_{ad} = J_x~~\sqrt{k_1^2 + 2 \alpha J_y/k_2} + k_2 J_y
\label{adH}
\end{equation}
Note that semiclassical quantisation of $H_{ad}$, which is also
exact in this case, gives for the ground state energy,
\begin{equation}
E_{ad} = \frac{1}{2} \sqrt{ k_1^2 + \frac{\alpha}{k_2} } + \frac{k_2}{2}
\end{equation}
The energy estimated by this formula is in good agreement with the
computed energies for small $\alpha$.
Since we are interested only
in the eigenstate, the last term in Eq. (\ref{adH}) can be omitted since
it only serves to shift the energy scale.
The ground state of $H_{ad}$ is,
\begin{equation}
\psi(x) = \left( \frac{\omega}{\pi} \right)^{1/4} \exp(- \omega x^2/2)
\label{gs_x}
\end{equation}
where $\omega = \sqrt{ k_1^2 + \frac{\alpha}{k_2} }$.
The ground state of the Hamiltonian in Eq. (\ref{Ham}), for $\alpha << 1$, in
the adiabatic approximation, can be written down in its characteristic form as,
$\Psi(x,y) = \phi(x;y) ~\psi(x)$, so that,
\begin{equation}
\Psi(x,y) = \frac{( \omega \omega_x)^{1/4} }{\sqrt{\pi}} ~e^{-\omega_x y^2/2} 
~e^{-\omega x^2/2}
\label{posgs}
\end{equation}
Note that $\Psi(x,y)$ is a coupled eigenstate; the coupling provided by $\omega_x$.
It is correctly normalised. For $\alpha=0$ it reduces to a two-dimensional
harmonic oscillator wavefunction. The price we pay in the adiabatic approximation
is that the eigenstate is not symmetric under $x \leftrightarrow y$ as required
by the potential.

\subsection{Momentum eigenstate}
We calculate the eigenstate in the momentum representation by taking Fourier
transform of the eigenstate given in Eq. (\ref{posgs}). We perform
the following integral,
\begin{equation}
\widetilde{\Psi}(p_x,p_y) = \frac{1}{2\pi} \int^{\infty}_{-\infty} \int^{\infty}_{-\infty} \Psi(x,y)
     e^{ip_x x} e^{ip_y y} ~~dx ~~dy
\label{ftrans}
\end{equation}
By a straightforward transformation of variables, we can reduce it to an
one-dimensional integral,
\begin{equation}
\widetilde{\Psi}(p_x,p_y) = \frac{\omega^{1/4}}{\sqrt{2}\pi} \int^{\infty}_{-\infty} dx
         \frac{e^{-\omega x^2/2} e^{-p_y^2/2\omega_x} e^{ip_x x}}{\omega_x^{1/4}}
\end{equation}
At this point, in order to be able to do the integration, 
we approximate the integrand for
$\alpha<< 1$. We take,
\begin{eqnarray}
\exp{\left( \frac{-p_y^2}{2\omega_x} \right)} & \backsimeq & \exp{\left( \frac{-p_y^2 (1- (\alpha/k_2^2)) x^2 }{2 k_2} \right)}  \nonumber \\
\omega_x^{-1/4} & \backsimeq & \frac{1}{k_2} \exp\left( -\frac{\alpha}{4 k_2^2} x^2 \right) 
\end{eqnarray}
Then, the integral reduces to a Fourier transform of a Gaussian,
\begin{equation}
\widetilde{\Psi}(p_x,p_y) = \frac{\omega^{1/4}}{\sqrt{2}\pi} \exp \left( \frac{-p_y^2}{2 k_2} \right) \int^{\infty}_{-\infty}
f(x) ~~e^{i p_x x} ~~dx
\end{equation}
where,
\begin{equation}
f(x) = \exp\left( \frac{-x^2}{2} \omega_p \right)
~~~~;~~~~ 
\omega_p = \beta - \frac{\alpha p_y^2}{k_2^3}
\end{equation}
and $\beta = \omega + \alpha/2 k_2^2$.
This integral can be performed and we obtain
the momentum eigenstate as,
\begin{equation}
\widetilde{\Psi}(p_x,p_y) = \frac{1}{\sqrt{\pi}}
\left( \frac{\omega}{k_2} \right)^{1/4}
\frac{ \exp( -\frac{p_y^2}{2k_2}) \exp(- \frac{p_x^2}{2 \omega_p}) }{\sqrt{\omega_p}}
\label{momgs}
\end{equation}
This eigenstate is also of the product form characteristic of the adiabatic
approximation,
$\widetilde{\Psi}(p_x,p_y) = \chi_1(p_x;p_y) \chi_2(p_y)$, where $\chi_1$ and
$\chi_2$ represent the eigenstates of fast and slow degrees of freedom.
In contrast to the position eigenstate in Eq. (\ref{posgs}), the ground state in
momentum representation is normalised only to $O(\alpha)$. This is due to
the approximations that were done in the course of obtaining the Fourier
transform. As a limiting case, for $\alpha=0$, it reproduces the 2D momentum
eigenstate.

\section{Ground state entropy}
\subsection{Position Entropy}
Now it is possible to evaluate the information entropies with the eigenstates 
in Eqs. (\ref{posgs},\ref{momgs}). The position density, from Eq.(\ref{posgs}),
is given by,
\begin{equation}
\rho(x,y) = |\Psi(x,y)|^2 = \frac{\sqrt{\omega \omega_x}}{\pi} e^{-\omega_x y^2}
e^{-\omega x^2}
\end{equation}
Then, the information entropy can be evaluated as,
\begin{eqnarray}
S_{q}(\alpha,k_1,k_2) & = & - \int^{\infty}_{-\infty} \int^{\infty}_{-\infty} 
\rho(x,y) \ln \rho(x,y) ~dx~dy  \nonumber \\
      & = & 1 + \ln\pi - \ln\sqrt{\omega} - \frac{I_1}{\sqrt{\pi}} 
\label{ent_int}
\end{eqnarray}
where $I_1$ is the integral given by,
\begin{equation}
I_1 = \int^{\infty}_{-\infty} e^{-x^2} ~\ln\left(k_2 + \frac{\alpha}{\omega k_2^2} x^2 \right) ~dx
\end{equation}
Though this integral can be exactly done, we expand the
integrand to first order
in $\alpha$ and arrive at the following expression for the entropy valid
for $\alpha << 1$,
\begin{equation}
S_{q}(\alpha,k_1,k_2) = 1 + \ln\pi - \ln\sqrt{k_1 k_2} - 
\frac{\alpha}{4} \frac{k_1+k_2}{k_1^2 k_2^2} + O(\alpha^2)
\label{entx}
\end{equation}
The spatial entropy decreases linearly with $\alpha$.
Setting $\alpha =0$, we recover the correct 2-dimensional harmonic oscillator
entropy.  In Fig \ref{ent_alpha}(b)  we show the spatial
entropy ($k_1=k_2=1$) as a function of $\alpha$.
For comparison, we numerically estimate the entropies by 
diagonalising the Hamiltonian in Eq. (\ref{Ham}) in harmonic oscillator basis set and a
subsequent entropy calculation. There is a good
agreement between the numerical results (dots) obtained without any approximation and
the analytical formula (solid line).
A linear fit to the numerical entropies gives a slope of
-0.4533, not far from the theoretical result -0.5. The fitted slope
closely approximates -0.5 as $\alpha \to 0$.

\subsection{Momentum Entropy}
\begin{figure}
\includegraphics[height=5cm]{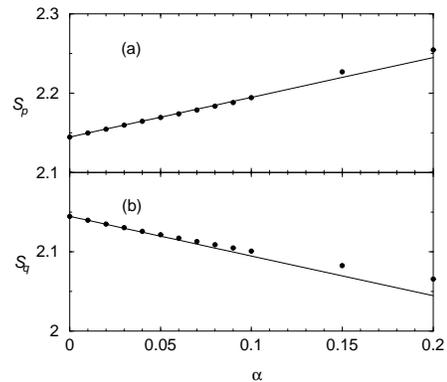}
\caption{Entropy for the ground state of the Hamiltonian in Eq. (\ref{Ham})
with $k_1=k_2=1$ as a function of $\alpha$.
(a) Momentum entropy and (b) spatial entropy. Dots are numerical values
of information entropies obtained without any approximation. Continuous
lines are theoretical curves of Eqs. (\ref{entx},\ref{entp})}
\label{ent_alpha}
\end{figure}

The momentum density, $\widetilde{\rho}(p_x,p_y)$, 
from Eq.(\ref{momgs}), is given by,
\begin{equation}
|\widetilde{\Psi}(p_x,p_y)|^2 = \frac{1}{\pi}
\sqrt{\frac{\omega}{k_2}}~
\frac{ \exp\left(-\frac{p_y^2}{k_2}\right) \exp\left(-\frac{p_x^2}{\omega_p}\right)}{\omega_p}
\end{equation}
Once again, an entropy integral similar to Eq.(\ref{ent_int}) can be performed
by Taylor expanding the denominator in the integrand to $O(\alpha)$.
The final, rather cumbersome, expression turns out to be,
\begin{equation}
S_p(\alpha,k_1,k_2) = \sqrt{\frac{\omega/\beta}{(1-c)}} ~\left( \frac{1}{2}
-\ln\left(\frac{\sqrt{\omega/k_2}}{\pi \beta}\right) +
\frac{(1-2 c)}{2(1-c)} \right)
\end{equation}
where we use the shorthand, $c = \alpha/2\beta k_2^2$.
As a limiting case, we get for $\alpha=0$, the correct momentum
entropy of 2D harmonic oscillator. To obtain an explicit expression, we
expand to $O(\alpha)$ all the terms in the above equation. The 
result valid for $\alpha<<1$, is,
\begin{equation}
S_p(\alpha,k_1,k_2) = 1+ \ln\pi + \ln\sqrt{k_1 k_2} + 
\frac{\alpha}{4} \frac{k_1+k_2}{k_1^2 k_2^2} + O(\alpha^2)
\label{entp}
\end{equation}
This expression shows that the momentum entropy increases linearly
with $\alpha$. This
is evident from Fig \ref{ent_alpha}(a) in which a good agreement is
seen between the numerical and the analytical result. A linear fit
gives the slope to be 0.4765, close to the expected value 0.5.

\begin{figure}
\includegraphics[height=5cm]{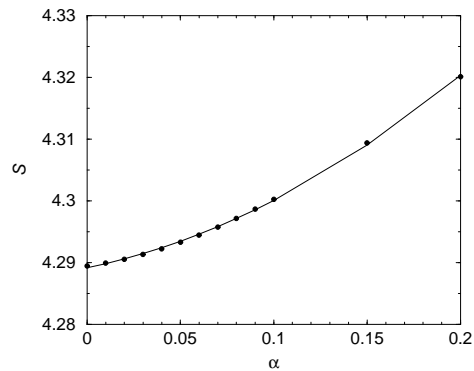}
\caption{Entropy sum for the ground state of the Hamiltonian in Eq. (\ref{Ham})
with $k_1=k_2=1$ as a function of $\alpha$.
Dots are numerical values and continuous line is the quadratic fit to
the numerical data.}
\label{entsum}
\end{figure}

\subsection{Entropic sum}
From the entropy expressions in Eqs. (\ref{entx},\ref{entp}), we obtain
the sum of position and momentum entropies as,
\begin{equation}
S(\alpha) = S_q + S_p = 2~(1+ \ln\pi) + O(\alpha^2)
\end{equation}
We point some salient features of this result. To $O(\alpha)$, the
entropic sum is invariant under change of $\alpha$,
but definitely satisfies the uncertainty inequality in Eq. (\ref{bbmeq}).
As expected, due to the scale invariance, it is also independent of 
the parameters $k_1$ and $k_2$. However, as revealed by numerical
computations in Fig \ref{entsum}, the significant contribution to
entropic sum comes from terms of $O(\alpha^2)$, which could not be
determined from the present method.
The important result is that, for
$\alpha <<1$, the entropies and their sum depend only on the
parameter $\alpha$, infact on terms of $O(\alpha^2)$ or higher.
Since our method does not yield information on the quadratic and higher
order $\alpha$ terms, we perform a fit to the numerical entropic sum. The
equation of best fit, for the range of $\alpha$ considered,
gives $S = 4.289 + 0.063 \alpha + 0.466 \alpha^2 $.
As expected, the fitted equation shows that the linear term in $\alpha$ 
carries less weight and the intercept is a good approximation to $2(1+ \ln\pi)$.
The ground state saturates the entropic sum only for $\alpha=0$.

The role of $\alpha$, primarily, is to change the
shape of the potential which leads to other qualitative changes in the
classical dynamics of the system. As $\alpha$ increases, the potential
develops 'channels' (see Fig. \ref{potential}) but the ground state does 
not occupy the channels.
The potential restricts the ground state to remain within the
central region and not expand into the channels of the potential. 
Hence the spatial entropy decreases with increase in 
$\alpha$ indicating stronger localisation of the 'particle' near the origin
of the potential.
Correspondingly, the momentum entropy increases.
But they do not cancel one another exactly, and the increase in
entropic sum is of the order $O(\alpha^2)$. Thus, it is a rather
slow but definite increase, as evident from Fig. \ref{entsum}.

\section{Large Parameter}
We note that the limit $\alpha \to \infty$ is a singular
limit for the Hamiltonian in Eq. (\ref{Ham}). It does not
seem feasible to extend the present method to this limit.
Hence we look at $\alpha >>1$, but not $\alpha \to \infty$.
\subsection{Position Eigenstate}
For $\alpha>>1$, we cannot Taylor expand as done in Eq. (\ref{app_omega}).
Hence, now the adiabatic Hamiltonian will be,
\begin{equation}
H_{ad} = \frac{p_x^2}{2} + \frac{k_1^2 x^2}{2} + \Omega_x J_y
\end{equation}
Presently, exact solutions are not known for this problem.
To go further, we quantise the $y$ degree of freedom exactly
and treat the $x$ degree of freedom using the variational method.
This leads to,
\begin{equation}
H_{ad} = \frac{p_x^2}{2} + \frac{k_1^2 x^2}{2} + \frac{1}{2} 
\sqrt{k_2^2 + 2 \alpha x^2}
\end{equation}
where $\Omega_x$ is substituted from Eq.(\ref{full_omega}).
Now, we will use variational method to obtain the ground state.
The trial wavefunction of the form,
\begin{equation}
\psi_{tr}(x) = \left( \frac{b}{\pi} \right)^{1/4} \exp(- b x^2/2)
\label{trial}
\end{equation}
is assumed.
Note that the trial wavefunction is similar to the adiabatic
ground state in Eq. (\ref{gs_x}), except for $\omega$ being
replaced by the variational parameter $b$, to be determined.
Since we are applying the standard variational technique as
expounded in many quantum physics books, we refer the  reader
to, say  \cite{tan}, for details of the variational method.
We first determine the energy functional,
\begin{eqnarray}
E_{v} & = & \langle \psi_{tr}(x) \mid H_{ad} \mid \psi_{tr}(x) \rangle \\
        & = & \frac{b}{4} + \frac{k_1^2}{4 b} + \sqrt{\frac{\alpha}{2b}} 
             \;\; U(-1/2,0,b k_2^2/2 \alpha)
\end{eqnarray}
where $U(,,)$ is the confluent Hypergeometric function.
Now, we optimise $E_{v}$ to obtain $b$. Hence,
\begin{eqnarray}
\frac{dE_{v}}{db} & = & \frac{1}{4} - \frac{k_1^2}{4 b^2} + 
 \frac{e^{b k_2^2/4\alpha} \; k_2^2 \; K_0(b k_2^2/4 \alpha)}{4 \sqrt{2b\pi\alpha}} \nonumber \\
  & & -\frac{\sqrt{\alpha}}{2\sqrt{2}b^{3/2}} \; U(-1/2,0,b k_2^2/2 \alpha)
\end{eqnarray}
where $K_0(.)$ is the modified Bessel function of second kind.

Now, we solve for $b$ by setting $dE_{v}/db = 0$. Once again, solving
for $b$ in  a general case does not seem feasible. However, in the
limit $\alpha >>1$, we use the asymptotic forms as $z = (b k_2^2/2 \alpha) \to 0$
\cite{asymp},
\begin{eqnarray}
U\left(\frac{-1}{2}, 0, z \right) & \thicksim & \frac{1}{\sqrt{\pi}} \left( 1 + \frac{z}{2} \left(
       \ln z + \Psi(\frac{1}{2}) + 2 \gamma -1 \right) \right) \nonumber \\
K_0(z/2) & \thicksim & - \ln(z/2)
\end{eqnarray}
where $\gamma=0.5772...$ is the Euler's constant and $\Psi(.)$ is the
digamma function.
We use these asymptotic forms and ignore terms that contain $\alpha$
in the denominator. Then, we obtain,
\begin{equation}
\frac{dE_{v}}{db} \simeq \frac{1}{4} - \frac{k_1^2}{4 b^2} - 
\frac{\sqrt{\alpha}}{2 \sqrt{2 \pi} b^{3/2}} = 0
\end{equation}
If the terms are rearranged, this turns out to be a quartic algebraic
equation in $b$.
\begin{equation}
b^4 - 2 k_1^2 b^2 - \frac{2 \alpha b}{\pi} + k_1^4 = 0
\label{beq}
\end{equation}
This can be exactly solved for $b$ but leads to rather complicated terms.
As an aside, we also point out that using asymptotic forms for $U(.)$ and
$K_0(.)$ as $z \to \infty$, i.e, $\alpha \to 0$, we can recover the
position eigenstate obtained in Eq. (\ref{posgs}).

From this point, we specialise to our case $k_1 = k_2 =1$. Then,
for $\alpha >>1$, the second term in Eq. (\ref{beq}) can be neglected
since $b$ increases monotonously with increasing $\alpha$.
Now, setting $dE_{v}/db = 0$, we obtain for $\alpha >>1$,
\begin{equation}
b \approx \left( \frac{2\alpha}{\pi} \right)^{1/3}
\end{equation}
In general, $b$ is dependent on $k_1$ and $k_2$ as evident from the
exact solution to Eq. (\ref{beq}) but we have taken $k_1=k_2=1$.
Plugging in this $b$ in Eq. (\ref{trial}), we obtain the adiabatic
ground state valid for $\alpha >>1$ as,
\begin{equation}
\Psi(x,y) = \frac{( b \Omega_x)^{1/4} }{\sqrt{\pi}} ~e^{-\Omega_x y^2/2}
~e^{-b x^2/2}
\label{gs_lalp}
\end{equation}
where $\Omega_x = \sqrt{1+2 \alpha x^2}$ since we have taken $k_1=k_2=1$
to make the analysis tractable. It is correctly normalised
but does not give the correct limit for $\alpha=0$. It does not possess
the symmetry $x \leftrightarrow y$.

\subsubsection{Position entropy for $\alpha >>1$}
The probability density in position representation is,
\begin{equation}
\rho(x,y) = |\Psi(x,y)|^2 = \frac{\sqrt{b \Omega_x}}{\pi} e^{-\Omega_x y^2}
e^{-b x^2}
\end{equation}
From this, the information entropy with $k_1=k_2=1$ can be evaluated 
(see Eq. (\ref{ent_int}) as,
\begin{equation}
S_{q}(\alpha) = 1 + \ln\pi - \ln\sqrt{b} - \frac{I_2}{\sqrt{\pi}}
\label{xent_large}
\end{equation}
where $I_2$ is given by,
\begin{eqnarray}
I_2 & = & \int^{\infty}_{-\infty} e^{-x^2} ~\ln \left( 1 + \frac{2\alpha}{b} x^2 \right) ~dx \nonumber \\
    & = & \sqrt{\pi} \left[ \gamma + \pi~\mbox{erfi}\left( \sqrt{\frac{b}{2\alpha}} \right) +
    \ln \left( \frac{\alpha}{2b} \right) -  \right.      \nonumber \\
    &   & \left. \frac{b}{\alpha} {}^{}_{\phantom{1}2}F_2 \left(1,1 ; \frac{3}{2},2 ; \frac{b}{2\alpha} \right) \right]
\label{I2}
\end{eqnarray}
and $\mbox{erfi}(.)$ is the imaginary error function and ${}^{}_{\phantom{1}2}F_2(.)$ is the
generalised hypergeometric function. In this form, the information entropy in Eq. (\ref{xent_large})
is not particularly illuminating. For $\alpha >>1$, the contributions from 2nd and 4th
term in Eq. (\ref{I2}) is negligible. Thus, we obtain a simplified form for the
spatial entropy,
\begin{equation}
S_{q}(\alpha) = 1 + \ln\pi - \ln\sqrt{b} + \frac{\gamma}{4} - \frac{1}{4}\ln \left( \frac{\alpha}{2b} \right)
\end{equation}
Now, substituting for $b$ from Eq. (\ref{beq}), we obtain,
\begin{equation}
S_{q}(\alpha) = 1 + \ln\pi + \frac{\gamma}{4} - \frac{1}{3}\ln\left(\frac{\pi}{2}\right) - \frac{1}{3} \ln\alpha
\label{ent_theory_large}
\end{equation}

Thus, for large $\alpha$, the ground state entropy falls logarithmically with
a slope 1/3. This is verified numerically through exact calculations.
In Fig. \ref{ent_large}(b), we show the position entropy
plotted as a function of $\ln(\alpha)$. A linear regression gives the best fit line
as $ S_{q} = 2.1368 - 0.3033 \ln\alpha$, verifying the approximate theoretical slope in
Eq. (\ref{ent_theory_large}). The intercept is approximately $1+\ln\pi$, the
entropy of the unperturbed oscillator, but we
stress that we cannot recover $\alpha=0$ result from Eq. (\ref{ent_theory_large}).
We also point out a systematic difference seen in Fig. \ref{ent_large}(a,b) 
between the theoretical estimate of
entropy (crosses) given by Eq. (\ref{ent_theory_large}) and the numerical
result (circles).
This is due to the adiabatic approximation
becoming less accurate near the origin of the potential and this is
showing up in the results.

\subsection{Momentum eigenstate and entropy}
\begin{figure}
\includegraphics[height=5cm]{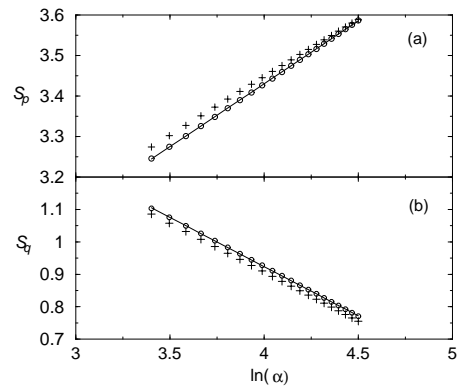}
\caption{(a) Entropy in momentum representation and in (b) spatial
representation for the Hamiltonian
in Eq. (\ref{Ham})
with $k_1=k_2=1$ as a function of $\alpha$
ranging from 30 to 90.
Dots are the numerical entropies and solid line is the best fit to
the numerical data. Crosses represent the theoretical
result based on the adiabatic approach.}
\label{ent_large}
\end{figure}

We obtain adiabatic ground state in the momentum representation for $\alpha >>1$
by Fourier transforming the position eigenstate in Eq. (\ref{gs_lalp}).
We need to perform the integral in Eq. (\ref{ftrans}) using the
ground state in Eq. (\ref{gs_lalp}). Once again,
by transforming variables, we can reduce it to a one-dimensional integral,
\begin{equation}
\widetilde{\Psi}(p_x,p_y) = \frac{b^{1/4}}{\sqrt{2}\pi} \int^{\infty}_{-\infty} dx
         \frac{e^{-b x^2/2} e^{-p_y^2/2\Omega_x} e^{ip_x x}}{\Omega_x^{1/4}}
\label{mom_int}
\end{equation}
where $\Omega_x$ is given by Eq. (\ref{full_omega}) and $b$ is the variational
parameter determined in Eq. (\ref{beq}). This integral could only be performed
numerically. Hence, we first determine the adiabatic momentum eigenstate by numerically
integrating Eq. (\ref{mom_int}) from which the entropies are computed.
The entropies obtained numerically by the adiabatic approach, shown as
crosses in Fig \ref{ent_large}(a), and those
obtained numerically without any approximation (circles in Fig \ref{ent_large}(a))
are in fair agreement with each other for the range of $\alpha$ considered in
this work. A linear regression gives the line of best fit,
\begin{equation}
S_p(\alpha) = 2.1754 + 0.3134 \ln\alpha
\label{bestfit2}
\end{equation}
Momentum entropy increases logarithmically with $\alpha$.
We notice from our numerical calculations, exact as well as the one based
on adiabatic approach, that the slope in the equation of best fit for
the momentum entropy is consistently
higher than the one for the position entropy. This difference accounts for
the behaviour of the entropic sum presented in the next section.

\subsection{Entropic sum for large $\alpha$}
\begin{figure}
\includegraphics[height=5cm]{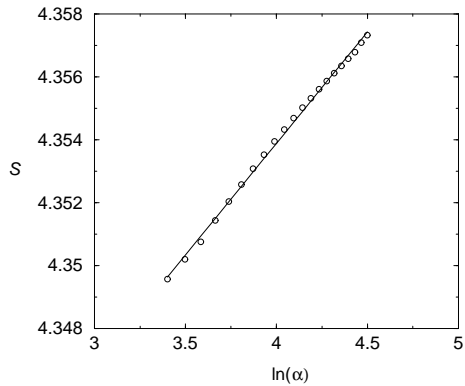}
\caption{(a) Entropy sum for the ground state of the Hamiltonian
in Eq. (\ref{Ham})
with $k_1=k_2=1$ as a function of $\alpha$.
Circles are numerical values and solid line is the best fit to
the numerical data as given by the Eq. (\ref{bestfit3}).}
\label{entsum_large}
\end{figure}

Finally, we put together the results to look at the entropic sum for large $\alpha$.
The Fig \ref{entsum_large} shows the entropic sum for the ground state 
for a range of parameters
$\alpha= 30,....90$. The best fit line (solid line) is given by,
\begin{equation}
S = 4.3249 + 0.00708 \ln\alpha
\label{bestfit3}
\end{equation}
Firstly, it satisfies the entropic uncertainty inequality in
Eq. (\ref{bbmeq}).
The positive slope in this equation shows that the entropic sum
increases with the parameter but rather slowly. It must also be pointed
out that
this slope in Eq. (\ref{bestfit3}) corresponds closely to
the difference in slopes corresponding to the best fit equations for position
and momentum entropy.

The functional form of Eq. (\ref{bestfit3}) is quite analogous to
the entropic sum in Eq. (\ref{geq}) conjectured for atoms, clusters and 
nuclei \cite{gad1,amnc}
on the basis of theoretical arguments and numerical results.
Such logarithmic
increase in entropic sum is noted earlier for 1D power-law potentials
in the semiclassical limit \cite{deh}.
Thus, this result
provides an approximate (semi-)thereotical approach to see the emergence of
the functional form in Eq. (\ref{bestfit3}) in a simple coupled system.

\section{Highly Excited States}

In this section, we look at the grey areas that are not yet
clearly understood.  We briefly discuss the EUR for the highly excited states.
Most eigenstates of chaotic systems in the semiclassical limit
are irregular or random looking states that could be modelled by random
matrix theory \cite{rmt}. Such random matrix averages represent a particular limit
at which the system's entropy becomes independent of the nonlinearity parameter.
Hence we focus
our attention on the highly excited 'regular' states
which are characterised by $N$ quanta of excitation in one degree of freedom
and 0 in the other, often called the Born-Oppenheimer type of
states \cite{prange,adia}.
Thus they could be approximately labelled by
the quantum number pair $(N,0; \alpha=\alpha_0)$, where $N>>1$ \cite{mss1}.
For these states, we will numerically explore the variation of entropies
as a function of $\alpha$.
In a sense, the entropies of highly excited states have
already been reported before in a different context \cite{mss}.
Firstly, we choose a
particular eigenstate characterised by $(N,0;\alpha_0)$ , in our case (110,0;0.0), and
then follow the eigenstate as a function of $\alpha$.
The calculations are also somewhat cumbersome and hence we sample only
a small parametric window.

\begin{figure}
\includegraphics[height=5cm]{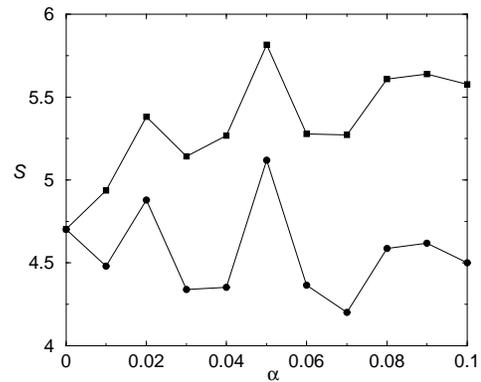}
\caption{Spatial (circle) and momentum (square) entropy for 
a highly excited state characterised by approximate quantum number
pair (110,0) with $k_1=k_2=1$ as a function of $\alpha$.}
\label{ent_110}
\end{figure}

The results, obtained from exact numerical basis-set diagonalisation
and then a subsequent entropy calculation, shown in the Fig. \ref{ent_110}
reveal that the spatial and momentum entropies display
similar trend as a function of $\alpha$. This is qualitatively
different from that of the ground state. From the figure, it is also
clear that the variation with $\alpha$ is not monotonous. The entropic
sum evidently obeys the inequality in Eq. (\ref{bbmeq}).
It is known that for Hamiltonian in Eq. (\ref{Ham}) there are
certain special periodic orbits which are responsible for supporting
regular states in the system \cite{anc}. The probability density structures
are related to the classical phase space structures \cite{bebo} and hence the
information entropy is also intimately connected to the properties of
the local classical phase space structures that support such regular
states. Thus the entropy is modulated by the qualitative nature
of local classical dynamics as a function of $\alpha$ \cite{mss1}.

\section{Conclusions and Discussions}

This work can be broadly divided into two parts. Focussing
on the ground state of a coupled system, firstly, we show,
within the framework of adiabatic theory,
that spatial and momentum eigenstates can be constructed.
Secondly, using the adiabatic eigenstates we obtain approximate
results for the entropic uncertainty relations.
For the ground state, we show that the spatial entropy
decreases as a function of the coupling parameter $\alpha$
in the system while the momentum entropy increases. However, the
entropic sum increases with the coupling parameter $\alpha$ and gets
saturated only for $alpha=0$. This is reminiscent
of the results reported numerically for the entropic sums of atoms,
molecules and clusters \cite{amnc}.
Thus, this work provides an approach to obtain analytical
results for the information entropies and their sums, for
{\it coupled} nonlinear systems as a function of coupling parameter.

However, for highly excited states, the spatial and momentum entropy
qualitatively look similar and seem to be related to the qualitative
nature of local classical dynamics. In general, the relation between the
classical phase space structures and the quantum eigenstates is not
completely understood yet. In this context, one might mention
the semiclassical approaches due to Berry and Bogomolny \cite{bebo} based on
the Gutzwiller's Trace Formula. Thus, we only point out the intricacies
involved in the entropic sums for the highly excited states. In due
course, it might become possible to relate information entropy
to classical orbits through these approaches. There are already
empirical results pointing to this connection \cite{mss1}.

The adiabatic theory based approach presented here is not
without demerits. Strictly speaking, the adiabatic effects
take over when there is clear separation in the time-scales of motion
in the two modes that constitute the system.
Thus, in the case of Hamiltonian in Eq. (\ref{Ham}),
the potential develops channels (see Fig. \ref{potential})
for large values of $\alpha$ and thus facilitates adiabatic effects.
The adiabatic theory is particularly suitable if the probability
density develops structures within these channels, as it happens
for some of the highly excited states. Since the ground state
does not enter the channel for large $\alpha$, it is likely
to become less accurate for very large values of $\alpha$.
Further, this work leaves an interesting question unanswered.
What happens in the limit $\alpha = \infty$ ?  The present
method of treatment does not seem adequate to answer this question.
We hope the results here could stimulate research on more
rigorous approaches to entropic sums in a wide variety of coupled systems.

\end{document}